\documentclass[10pt,a4paper]{article}
\usepackage{graphicx}
\usepackage{tabularx}
\usepackage{multicol}
\usepackage{tabto}
\usepackage{url}
\usepackage{enumitem}
\usepackage{textcomp}
\usepackage{eurosym}
\usepackage{amsmath}
\usepackage{MnSymbol}
\usepackage{wasysym}
\usepackage{xcolor}
\usepackage[labelfont=bf]{caption}
\usepackage[yyyymmdd]{datetime}

\usepackage{flowchart}
\usetikzlibrary{shapes,arrows.meta,chains,external}
\newcommand{\cz}[1]{\textit{\textbf{#1}}}

\begin{document}

\bibliographystyle{plain}
\thispagestyle{empty}

\begin{center}
\large{\bf The Development of Central Bank Digital Currency in China: An Analysis}\\
\end{center}
\vspace{0.5em}
\begin{center}
\begin{minipage}[t][][t]{0.45\linewidth}
\begin{center}
{\large Geoffrey Goodell}\\
\vspace{0.5em}
{\large Centre for Blockchain Technologies}\\
\vspace{0.5em}
{\large University College London}\\
\vspace{0.5em}
{\large \texttt{g.goodell@ucl.ac.uk}}
\end{center}
\end{minipage}
\begin{minipage}[t][][t]{0.45\linewidth}
\begin{center}
{\large Hazem Danny Al Nakib}\\
\vspace{0.5em}
{\large Centre for Blockchain Technologies}\\
\vspace{0.5em}
{\large University College London}\\
\vspace{0.5em}
{\large \texttt{h.nakib@cs.ucl.ac.uk}}
\end{center}
\end{minipage}
\end{center}

\begin{center}
{\textit{This Version: \today}}\\
\end{center}
\vspace{4em}

\begin{abstract}

The People's Bank of China (PBOC) has launched an ambitious project to develop
a digital currency for use in domestic, retail transactions, and is, by far,
the most advanced globally in this regard.  In addition to involving a diverse
set of stakeholders, the PBOC established a set of fundamental principles,
including privacy, inclusiveness, and conservatism, and has articulated its
progress in a public document translated into English.  We maintain that
although both its first principles and its conclusions drawn from the research
conducted by the PBOC from 2014 to date are broadly reasonable and appropriate,
the PBOC has also missed some important considerations and entertained some
questionable assumptions, which many central banks around the world have also
done.  In this analysis, we consider the strengths and weaknesses of the
digital currency proposition articulated by the PBOC as it exists today, and we
propose one fundamental and specific change for the PBOC and other central
banks around the world: The architecture must accommodate privacy-preserving,
non-custodial wallets.  With this change and a related set of minor
adjustments, China has an opportunity to lead the world in the implementation
of a central bank digital currency (CBDC) solution that protects the authority
of the central bank to implement monetary policy, preserves the role of
public-sector and private-sector banking institutions, promotes the efficiency
of retail transactions and businesses, satisfies regulatory objectives, and
safeguards the human rights of retail consumers, including their privacy and
their right to participate in the economy.  We hope that the PBOC, and other
central banks around the world, will have the resolve and strength of purpose
to implement our proposed change and carry on with implementing a CBDC
architecture that serves the interests of its users.

\end{abstract}

\section{Introduction}

In July 2021, the People's Bank of China (PBOC) published a report detailing
its progress, which began in 2014, toward deploying a central bank digital
currency (CBDC) in China~\cite{pboc2021}, with over \$5.3 billion reported in
transactions to date through various pilot programs~\cite{bloomberg2021}.
The stated purpose of this report was to ``clarify the PBOC's position'' and to
``explain the background, objectives and visions, design framework and policy
considerations for the e-CNY system''~\cite{pboc2021}.  This effort to engage
the general public is laudable, and, to be clear, there is much to commend in
this effort, its central principles, and the general direction of its progress.
In this brief response, we characterise the salient features of the proposed
approach to CBDC and their applicability to the efforts underway to develop
CBDC worldwide.

We observe that a potentially conflicting set of requirements, which could
undermine the success of this project if left unresolved, arise from some
unfounded architectural assumptions about how institutionally-supported digital
currency must work.  Most importantly, the PBOC proposal treats wallets as
ontologically indistinguishable from accounts, a conflation made by many
central banks around the world.  As a result, not only does the system
introduce new costs and risks related to the institutional management of such
wallets, but it compels retail users to relinquish control of their assets and
exposes them to the risk and harm of unwelcome profiling. This is particularly important
because the general tenor of the discussion and debate surrounding CBDC
worldwide, including by governments and central banks, has viewed CBDC as a
general and complete substitute for physical cash.

This article is organised as follows.  In the next section, we characterise the
design goals and requirements identified by the PBOC.  Next, we evaluate the
various assumptions implicit to the approach.  Then, we specify a specific set
of recommendations based upon our analysis.  Finally, we conclude with some
remarks about the implications of this proposal for CBDC efforts currently
underway within the international community.

\section{Requirements}

Arguing from first principles, the authors at the PBOC make a case for a set of
desiderata for the e-CNY as a domestic retail payment system.  This fact alone
is significant.  Some of the prevailing proposals for CBDC focus on addressing
the inefficiencies of correspondent banking in a cross-border
context~\cite{ripple2015}.  This problem is certainly real and worth solving,
although this perspective does not address the real danger of vanishing
infrastructure to support cash, as identified by the PBOC research and
experienced directly by countries such as Sweden~\cite{sennero2019}.  Other
proposals seem to assume that CBDC would primarily support or enhance the
wholesale payment infrastructure~\cite{r3-cbdc}.  However, the requirements of
wholesale and retail users of cash are quite different; in particular, human
consumers have human rights.  Therefore, the PBOC is right to acknowledge that
we need a purpose-built solution.  The PBOC identified the following
desiderata, among others, as core requirements for the design of
e-CNY~\cite{pboc2021}:

\begin{itemize}[label={\small $\blacksquare$}]

\item \cz{Inclusiveness.}  The PBOC acknowledged financial inclusion as a
primary objective for CBDC, and they are right to do so.  The authors
specifically noted the decline of cash as a motivating factor for this
objective, and they acknowledged technological progress as a contributing
factor to the decline of cash.  As the variable revenues associated with
operating a cash infrastructure fall below the fixed costs, maintaining the
cash infrastructure becomes untenable.  However, although in principle anyone
can acquire and possess cash, not everyone is in the position to acquire modern
digital money, much less possess it.  Bank deposits require banking
relationships, and digital transactions require access to network connectivity,
electricity, and the skills necessary to make good use of modern technology.
The secular decline of cash risks jeopardising the ability for certain persons
(for example, the impoverished, the elderly, and the disabled) and even entire
communities (for example, the most remote) to fully participate in the modern
economy.  We argue that the PBOC is right to implicitly define financial
inclusion in terms of ability to interact with the economy, rather than in
terms of access to banking services and credit facilities, as others, including
the World Bank~\cite{world-bank}, as well as several central banks and
governments.  CBDC might hope to provide a public option, and its many
possible forms might hope to reach even the most disadvantaged groups.

At the same time, we must be cautious not to treat CBDC as a panacea for
financial inclusion, as that might easily lead to considering CBDC as a direct
and complete substitute for cash in circulation and to no longer include
providing access to cash among the paramount policy objectives of a central
bank, namely to ensure that all persons have access to the economy. The authors
of the report have clearly stated that the role of the central bank includes
the responsibility to ensure that the public has direct access to cash and that
this responsibility would also apply in the case of CBDC. It is unclear whether
the PBOC maintains this in reference to the developments in the digital age
which it references, the developments in technology, or the decrease in the use
of cash.  In any case, it is a fundamental part of any discussion surrounding
the development of a CBDC architecture that many central banks have overlooked.
Public, non-discriminatory access to money ought to be an objective of every
central bank globally, and the decrease in the use of cash in some locations by
certain demographics does not imply that the need for cash will eventually end
or that somehow, the deployment of CBDC will guarantee financial inclusion
\textit{simplicitur}.

\item \cz{Privacy.}  It should come as no surprise that the PBOC expressed
concern about the privacy of retail consumers.  In February 2020, the Shanghai
and Hong Kong Stock Exchange suspended the much-anticipated IPO of Ant
Financial, following the expression of ``concern about excess leverage and
abuse of consumer data'' by the government of China~\cite{goldman2021}, an
alleged practice that has surfaced in other jurisdictions, with other
organisations and other governments.  In effect, China acknowledged the dangers
of allowing technology platforms to accumulate, aggregate, and analyse consumer
data at scale.  In particular, the risk that CBDC could become a powerful tool
for predicting and influencing the behaviour of populations through
surveillance is well-acknowledged~\cite{gladstein2021}, both by government and
by private-sector intermediaries and providers of infrastructure solutions, and
such an outcome is explicitly deemed undesirable by the PBOC, at least for
retail transactions of ``small value'', for which the authors specified
untraceable anonymity as a requirement~\cite{pboc2021}. For these reasons, the
authors of the PBOC report emphasise that the e-CNY ``can be transferred
without relying on a bank account'' and that it supports ``off-line''
transactions, referring to their approach as ``managed anonymity''.

\item \cz{Safety.} The authors indicated that CBDC must be safe for retail
consumers to use.  The PBOC report implicitly demonstrates that safety has at
least two dimensions.  The first dimension involves the extent to which retail
consumers should consider CBDC to be a risk-free asset, which implies that CBDC
must not involve counterparty risk.  CBDC, which is issued by a central bank,
is risk-free by definition and constitutes a liability on the balance sheet of
the central bank, as a debt owed by the central bank to the users of the CBDC,
similar to banknotes and coins.  However, private money, including without
distinction stablecoins such as ``Tether'' and commercially-issued digital
currency such as Facebook's renamed ``Diem'' (formerly
``Libra''~\cite{khan2019}), intrinsically incur counterparty risk, among a
variety of other risks and costs, as the guarantors of their value could
fail~\cite{gorton2021}.  The second dimension of safety involves the extent to
which CBDC must remain available for retail consumers to access and use, must
remain stable in value, and must remain widely accepted by retail merchants,
particularly during periods of economic distress. This view of safety could be
taken to imply that CBDC must provide a safer haven to hold one's wealth than
commercial bank deposits.  If so, then it could mark a significant shift away
from traditional responses in periods of economic distress.  Whether or not
this is systemically appropriate is beyond the scope of this paper. Given the
repeated use of the terms ``safe'' and ``safety'', and considering the
requirements for privacy, it is fair to conclude that safety implies that
retail users must be free to spend CBDC without fear that information about
their private transactions could be used against them in any manner.  Our
understanding of the PBOC report supports the view that the PBOC agrees with
this, at least for 'small value' transactions.

\item \cz{Efficiency.} The authors expressed concern about not only the
implications of the decline of cash for financial inclusion but also the
economic motivations for the decline of cash.  In particular, the authors cited
high costs of cash management, including ``banknote design, minting and
printing, transportation, deposit and withdrawal, identification, processing,
reflow, destruction, [and] counterfeit protection''~\cite{pboc2021}.  Digital
technology implicitly promises to reduce these costs, and mobile and card
payments have substituted for cash in much of China and indeed much of the
world.  A similar degree of efficiency would be required of any public payment
mechanism for that mechanism to be suitable, and competitive, in the digital
economy. This relates to the speed of transacting and settlement, the user
experience, and various the costs and risks involved in using a public payment
mechanism in contract to private-sector solutions, payment rails, and gateways.
Much of the approach to CBDC within the international community centres upon
finding the appropriate mechanism for a public payment infrastructure to
operate through private-sector participants and existing private sector
solutions, and approach initially pioneered by the PBOC, and which has also
been adopted in reports by the central banks of other countries, including the
Bank of England, its 2020 discussion paper~\cite{boe2020}.

\item \cz{Regulatory Compliance.}  The authors explicitly indicate that CBDC
would be implemented in a manner that would be compatible with compliance
requirements on money laundering and terrorist finance, which we interpret as a
reference to the FATF-GAFI Recommendations~\cite{fatf-recommendations}.  These
recommendations stipulate that custodians must require personally identifying
information for account holders.  We observe that the requirement for
collecting information about account holders is not necessarily incompatible
with the requirement for privacy that an account holder might have; for
example, today, an account holder can typically withdraw cash and spend it with
minimal risk of profiling.  However, as with the requirement for privacy, the
requirement for compliance also has important implications for the design of
the CBDC architecture.

\item \cz{Interoperability.} The authors identified several aspects of
interoperability as requirements for CBDC.  First, they noted standards
compliance, citing the need to establish international standards within
internationally agreed frameworks and mentioning the involvement of PBOC in
efforts to establish standards for digital fiat currency.  As of July 2021,
development of standards for digital fiat currency are still in early stages.
Although nascent efforts to develop international standards for digital
currency are underway within standards development organisations such as ISO,
CEN, ITU-T, and others, the international community and leading government
organisations have yet to rally behind any particular effort.  Second, the
authors argued that CBDC should be compatible with established financial
procedures, institutions, and regulatory processes.  When combined with the
requirement for efficiency, this requirement would seem to imply that it would
be preferable to leverage existing procedures, e.g. for clearing, settlement,
interactions with the central bank as part of the monetary base (i.e., a form
of M0), and regulatory compliance, rather than to establish new procedures for
CBDC.  Finally, the authors argue that it will be necessary to promote a
healthy ecosystem of CBDC-based services and solutions, including but not
limited to wallets.

\end{itemize}

We suggest that, taken individually, every one of these requirements is
reasonable and appropriate.  However, some of these requirements have
system-level implications for the design of CBDC, and some of these
implications have been overlooked by the authors and many central banks and
governments around the world.  For example, under certain assumptions, the
system-level implications of the requirements for inclusiveness and privacy
might conflict with the system-level implications of the requirements for
efficiency and regulatory compliance, in which case policymakers must undertake
a balancing act in the interpretation of these priorities to achieve the
optimal system architecture, with some considerations taking precedence over
others.  In the next section, we show that such a balancing act is not
necessary.  However, for a system to achieve all of the objectives, wallets
must be under the possession and control of the owners of the CBDC that they
contain.  Otherwise, inclusiveness, safety, and privacy will be significantly
curtailed, with a significant, deleterious effect on individuals, households,
and society.  Although some of the design features of our proposed
recommendations might seem to contradict certain requirements, such as
efficiency and regulatory compliance, this is actually not the case.  We argue
that it is possible to satisfy these requirements by reconsidering the
underlying assumptions and adjusting the desired architecture to different
assumptions instead.  In the next section, we show how this can be achieved.

\section{Assumptions}

The authors of the PBOC report base their proposed approach on several
assumptions, some of which are likely to be internally unsound. The assumptions
are common among central banks and governments worldwide and require attention.
In this section, we identify and characterise those assumptions.  We explain
why we believe that the assumptions are problematic and will raise issues that
the authors and other central banks have not adequately considered.  We also
explain how those assumptions might interfere with the development of a design
that meets the requirements outlined in the previous section and the optimal
balance between efficiency, regulatory compliance, inclusiveness, and privacy.

With regard to the balancing act, we do advocate for privacy, but we do not
believe that compliance implies sacrificing privacy or efficiency, safety or
inclusiveness.  It is fundamental understand what is meant by compliance: AML/KYC
regulations and FATF-GAFI recommendations apply to asset custodians, with the
purpose of preventing abuse of their privileged positions.  To achieve
compliance with these rules, custodians usually carry out identity checks on
their account holders.  We do not suggest changing such rules or procedures.
Instead, we suggest that the salient characteristic of digital currency is that
it can be possessed and controlled directly by its owners, outside custodial
accounts. That is, without the requirement that a third party rather than the
owner must possess the assets.  We imagine that those who exchange cash or bank
deposits for digital currency would usually be known by regulated money
services businesses (authorised or licensed businesses such as banks or
payments businesses).

\subsection{Wallets}

Our most significant concern arises in the use of the term ``wallet''.  What is
a wallet, exactly?  The authors of the report describe a wallet as something
that is provided, opened, and managed by ``authorised
operators''~\cite{pboc2021}.  These characteristics are nothing like the
characteristics of physical wallets, and this discordance might explain the
vagueness of the terminology chosen by the authors to describe their proposed
wallets as ``quasi-account-based'' and ``loosely coupled with bank accounts''.
In fact, the characteristics highlighted by the authors are about
\textit{relationships} between the owners of money and third-party gatekeepers.
More specifically, they are about relationships between retail consumers and
financial custodians.  In other words, they are about a means (software) by
which retail consumers can access their \textit{accounts}, which contain assets
that are ultimately under the control of a custodian.  This is, of course, a
significant departure from cash, which is directly possessed by its owners and
is under the direct control of its owners. It therefore becomes evident that
the authors' use of the term ``value-based'' to describe their CBDC system
cannot be disaggregated from their characterisation of it as variously
``account-based'', ``quasi-account-based'', and ``loosely coupled with bank
accounts''.

The terms ``account-based'', ``value-based'', and ``quasi-account-based'' were
used by the PBOC.  These terms are somewhat vague.  We interpret
``value-based'' to indicate that the account holds a specific quantity of
currency, as opposed to, for example, a credit score.  We interpret both
'account-based' and ``quasi-account-based'' to mean that an asset custodian is
maintaining a record that links multiple transactions to each other.  We
interpret 'account-based' to mean that the owner of the assets has a custodial
account, which would presumably comply with AML/KYC requirements for the
identification of the owner, and we interpret ``quasi-account-based'' as
pseudonymous, akin to a pre-paid debit card. A custodial account refers to
ownership and possession of the assets (in this case, the CBDC itself). For
example, banks are custodians of deposits. A custodial account entails that a
third party, not the owner, is in possession of the assets, and for a system to
require custodial accounts implies that its users simply do not have a choice in the
matter.

A close look shows that the use of the term ``wallet'' by the authors suggests
that transactions would indeed require the involvement of an authorised
gatekeeper. Therefore, what the PBOC appears to be describing is an
``account-based'' system wherein transactions are necessarily linked to
``accounts'' and not privately held software or hardware modules by the owners
of those assets, as defined in ISO 22739~\cite{iso-22739}. The main question
is: To transact digital currency, must a user implicitly ask permission from an
authorised business or other party that actually possesses the assets and
therefore serves as gatekeeper?  If so, then the system is similar to
commercial bank deposits.  If not, then it is similar to cash.  Whilst we agree
with the PBOC that it should be similar to cash, we do not agree that the
approach proposed in the article actually achieves this objective.

\subsubsection{Wallets as a replacement for cash}

We understand that the use of cash is declining in industrialised countries.
Part of the reason is the relative efficiency of digital payment channels.  The
result is that the variable revenues of operating a cash infrastructure are
falling below the fixed costs, and, as a result, the maintenance of cash
infrastructure is becoming increasingly untenable.

In the UK, for example, facilities to deposit and withdraw cash have become
less prevalent, debit cards and e-commerce (Internet) payments via custodial
accounts are capturing a larger share of payments, and brick-and-mortar
merchants and vendors are increasingly refusing to accept cash. The systematic
replacement of cash by intermediated forms of payment via custodial accounts is
harmful to consumers.  In particular, cash affords its users the following
benefits that modern retail payments do not:

\begin{itemize}[label={\small $\blacksquare$}]

\item\cz{Accessibility.} Nearly everyone can use it. The user interface for
cash is simple, it has built-in features for accessibility and security, and
its users of cash (banknotes and coins) do not require special registration,
contracts, bank accounts, network connectivity, or even electricity to use it
to make payments.

\item\cz{Non-Discrimination.} Because the value of cash is intrinsic,
everyone's money is as good as everyone else's. If one person gives cash to
another, the amount of value that one party relinquishes is exactly equal to
the amount of value that the other party gains, which is settled immediately on
receipt.  What cash can do is not specifically enabled or limited by the
identity of its bearer, nor of any party to the transaction.

\item\cz{Privacy.} Users of cash have no reason to fear that their
activities will be profiled on the basis of their transactions. Most cashless
payment methods leave behind a data trail that can be used to construct a
detailed history of an individual's habits, location, circumstances, and
psychology.

\item\cz{Ownership.} Cash is truly owned by its bearers. Users of cash know
that their transactions will succeed without the risk that a third party might
block them, in contrast to instruments that are ultimately under the control of
third parties and for which their users have only limited rights.

\end{itemize}

The systematic replacement of cash by intermediated forms of payment via
custodial accounts also undermines monetary sovereignty directly.  Funds held
as cash are the direct obligations of the central bank, whereas funds held in
custodial accounts are the obligations of private-sector banks or, potentially,
other potential financial services businesses that have access to central bank
accounts and reserves (depending on the jurisdiction).  Government deposit
insurance obscures but does not eliminate the difference between these two
forms of money from ordinary consumers.  When consumers conduct transactions
using the obligations of private-sector banks, the effectiveness of monetary
policy is undermined, both because multinational banks can transparently shift
the risk among currencies by taking advantage of their economies of scale and
because multinational banks can reduce their reliance upon the central bank of
the country in which their account holders conduct transactions, an important
point that is under-researched presently and must be considered when
redesigning and evaluating core economic mechanisms within society.

Additionally, stipulating that wallets must take the form of custodial
relationships raises some important questions about the custodians.
Governments of nations around the world, including China, have objected to the
abuse of consumer data by private-sector companies.  By forcing consumers to
use custodial accounts to hold and transact with digital currency, consumers
are exposed to these risks, \textit{mutatis mutandis}.  The question of how to
regulate such authorised operators also introduces challenges.  Establishing
the governance mechanisms and oversight that will be necessary to ensure fair
dealing and compliance with regulations, including not only regulations
concerning customer data but also concerning unfair dealing, discrimination,
and anti-competitive business practices, is no small task.

The authors of the PBOC report lightly touch on this final issue through
their discussion of ``managed anonymity'' whereby ``the PBOC sets up a firewall
for e-CNY-related information, and strictly implements information security and
privacy protocols, such as designating special personnel to manage information,
separating e-CNY from other businesses, applying a tiered authorization system,
putting in place checks and balances, and conducting internal
audits''~\cite{pboc2021}.  Presumably, the objective of such processes would be
to discourage ``arbitrary'' access to that information by ``authorized
operators'', which it identifies as primarily comprising ``commercial banks and
licensed non-bank payment institutions''~\cite{pboc2021} that circulate the
e-CNY to the public.  However, as the entities that the authors call
``wallets'' are essentially accounts, it is not obvious that the data
protection mechanisms recommended by the authors can achieve their stated
goals, and, given the system-level implications highlighted, the role of system
operators as arbiters of such mechanisms underscores the trade-off that is
inherently being made among the guiding principles.

\subsubsection{Non-custodial wallets}

Private, non-custodial wallets would allow the bearer of digital
currency to make a payment without linking the payment to any custodial
account; this feature is paramount to any CBDC design.  Specifically, it would
be possible for a bank to identify its customer when she withdraws digital
currency into her private wallet, just as it would identify her when she
withdraws cash.  The process of withdrawing digital currency would use
privacy-enhancing technology to strip out any characteristics that could be
used to link the digital currency to its owner.  Then, although the digital
currency held in her private wallet is not linked to her identity, the owner
would not be able to spend it except by remitting it to another regulated money
services business.  Typically, this would be an account with the payee's bank,
which would be obligated to identify the payee and associate the payee with the
transaction, although the payer would remain anonymous.  As with payments
involving cash, there is nothing to require that the time, location, or other
metadata associated with the withdrawal of CBDC from the payer's account into
her private wallet would be linked with the transaction wherein CBDC is
deposited into the payee's account.

This is what we mean when we say that there is no need to compromise either
privacy or compliance, or efficiency and inclusiveness, for that matter, in the
design of a CBDC system: The payer is subject to regulation at the time of
withdrawal, and the payee is subject to regulation at the time of the payment.
At the same time, non-custodial wallets offer the following benefits:

\begin{enumerate}

\item The withdrawal is unlinked from the payment.

\item The payer's spending habits remain private and cannot be observed without
the payer's explicit consent.

\item The payer avoids the risk of profiling.

\item The privacy is verifiable through the use of privacy-enhancing technology

\item The payer can trust mathematics rather than a trusted third party.

\item The payer is not forced to carry or use trusted hardware or
vendor-specific equipment.

\end{enumerate}

Moreover, this system would actually facilitate more regulatory oversight than
cash, because every payment must be processed by a regulated money services
businesses.  We hope that central banks would consider the benefits of
non-custodial wallets and endorse their implementation and deployment.

\subsection{The Use of CBDC}
\label{ss:use}

The authors of the PBOC report emphasise that CBDC would be part of the
monetary base (M0), not private-sector money.  This is a good thing, thus
ensuring that its users would hold risk-free assets.  Having said that, if
wallets are (by definition and per the previous section) accounts, then is this
just tantamount to requiring banks to segregate the CBDC of their depositors
rather than use fractional reserve banking to generate new money?  This would
seem bizarre, if we have understood correctly, both as a matter of policy, as
well as from the standpoint of a commercial business model to generate revenue.

The authors of the PBOC report contend that because CBDC is part of the
monetary base, it is a direct substitute for cash.  After all, it is a direct
obligation of the central bank and would allow its users to avoid
intermediation by existing or incumbent payment networks.  This is correct, as
is their assertion that a CBDC is primarily for retail payments. At the same
time, this does not answer all of the relevant questions, and we must ask how
it is a substitute, for what purpose, along which dimensions, and for whom. The
question of whether CBDC is a good substitute for cash and not a good
substitute for anything else is less obvious. We examine both aspects of the
argument, one at a time:

\begin{itemize}[label={\small $\blacksquare$}]

\item\cz{Is CBDC a good substitute for cash?}  First of all, what are the
motivating reasons and factors for people using cash for some transactions
today?  We can plainly identify many motivating reasons and factors, and not
having a bank account is only one of them.  Other factors include privacy, lack
of infrastructure, the economic cost of devices, and so on.  Some people live
in remote villages or transact in environments in which networked or electronic
payments are not an option.  CBDC would not be an adequate substitute for cash
in such cases, or in any of the cases where the motivating reason or factor
does not apply to CBDC in the same way that it does to cash.  For example,
desiderata for privacy, possession, or control as justifications for using cash
would not be satisfied by a system wherein a ``wallet'' essentially entails an
``account''.

\item\cz{Is CBDC a good substitute for other forms of money?}  In the
retail context, we limit our examination of this question to the other form of
money most familiar to retail consumers: commercial bank deposits.  If CBDC
were to take the form of non-custodial tokens, then it would indeed not be a
better store of value than bank deposits, as it would not be rehypothecated and
would not earn interest.  (Although mechanisms to simulate interest via
transfer payments to bearers could be implemented~\cite{goodell2020}, such
mechanisms would have implications for other guiding principles identified in
the PBOC report.)  However, implementing CBDC as non-custodial tokens would
allow owners of CBDC to make payments without involving custodial accounts, so
CBDC might indeed substitute for bank deposits, as cash does, in a limited and
specific capacity as a means of payment. Another possibility is that, from the
standpoint of financial risk and notwithstanding government insurance schemes,
CBDC might be safer than commercial bank deposits during periods of economic
distress.

\end{itemize}

CBDC is a direct obligation of the central bank, and could rightly be seen as a
form of cash that is upgraded for the digital economy.  We believe that it
would indeed substitute for cash in some cases, and if designed correctly,
could address concerns about privacy, accessibility, ownership, and
discrimination.  However, even in the best circumstances, it would not be a
perfect substitute across all the different motivating reasons and factors.
Some of these reasons and factors are critical to citizens, particularly the
vulnerable, and involve choices and decision-making that are essential for all
persons, and some of these reasons and factors are critical to the foundational
purpose of every central bank to ensure that persons within their purview have
efficient, ongoing, and sustainable access to the economy.  We also conclude
that although CBDC would not substitute for bank deposits as a store of value,
it could make an excellent substitute for bank deposits as a means of payment,
particularly for electronic payments in low-risk transactions that do not
require the payer to be traceable.

\subsection{Privacy and Managed Anonymity}

Of course, data protection is no substitute for privacy~\cite{nissenbaum2017}.
Users cannot trust what they cannot verify, and there is no way to verify that
information, once revealed, has not been copied, analysed, leaked, shared,
sold, stolen, or otherwise misused against the interests of the data subjects.
Some central banks around the world are making a mistake by conflating data
protection with privacy in the context of the design of CBDC architecture.

The authors of the PBOC report state clearly that counterparties should be
anonymous for ``small value'' transactions, and this policy is consistent both
with regulations and practices for low-risk transactions in other developed
countries throughout the world and with the human right to
privacy~\cite{un-human}.  We assert that the vast majority of transactions
conducted by most consumers are low-risk.  Therefore, to ensure that privacy is
not compromised by the de facto operating conventions of the new technology, we
assume that requirements for ``small value'' would be specified such that the
vast majority of retail transactions, and not some small subset, would fall
into that category.  More generally, although the term ``managed anonymity''
would seem to connote a prudent treatment of privacy rights, there are
important caveats that could limit the effectiveness of what the authors
propose, depending upon how it is implemented.

First, privacy is a public good~\cite{fairfield2015}, and the choice to not
reveal metadata is not really a choice that individual users can make for
themselves.  If most people are successfully incentivised to waive their rights
to privacy, then no one can be private, including those persons who did not
choose to waive their rights.  Second, allowing users to decline to furnish
personally identifying information when opening an account does not guarantee
that their transactions will remain private.  Consider the case of anonymous
accounts or quasi-accounts, which we understand as being comparable to pre-paid
debit cards.  To the extent that successive transactions with the same account
constitute linked attributes about a user, the user is pseudonymous and not
anonymous.

We believe that the salient question is how much privacy we want ordinary
persons to have, and with ``managed anonymity'', the answer is not much.
Inasmuch as transactions use accounts, which might be described as ``wallets''
(in contrast to non-custodial wallets), the successive transactions that
pseudonymous users make using such accounts would be linked to each other,
hence the author's reference to a ``loosely-coupled account linkage''.  So,
while a consumer could in principle use an anonymous account to make a single
payment, liquidating the entire account, most typical use cases would see
consumers acting as price-takers, so an exact match between the price of an
item and the value in an account would be unlikely.  The user of an account,
having completed a purchase of value less than the value of the account, would
then either use the account again for a subsequent transaction or forfeit the
difference between the agreed price and the value of the account.  As a result,
we might conclude that while malicious actors could use anonymous accounts to
achieve a reasonable level of anonymity, anonymous accounts would afford much
less privacy to ordinary, non-nefarious users.

Consumers have a right to conduct low-risk transactions with merchants,
vendors, and other providers of retail goods and services, without revealing
personal information that can be used to associate themselves with the
transaction.  For the avoidance of doubt, this includes any persistent
identifier associated with the consumer as well as any reference to another
transaction done by the consumer. Crucially,  it is possible to regulate
transactions without collecting or monitoring such personal information of the
consumer that can be used to associate a payer with a transaction.  In
particular, it is possible to collect information about the payee for
compliance with tax and AML/KYC regulations, along with relevant information
about the size, location, and nature of the transaction, while allowing the
payer to be anonymous.  As we have elaborated separately~\cite{goodell2021},
this requirement could be achieved with a CBDC system that permits transactions
only by regulated money services businesses and requires payees to be
account-holders or other parties who have fulfilled AML/KYC checks. We imagine
that even in such systems, there would be exceptions to the rule requiring
recipients to be identified, although these exceptions could be implemented
outside the payment mechanism itself.

We acknowledge that `perfect' anonymity is never really possible in practice,
as the size of an anonymity set is limited by the amount of cash in-flight at
any moment, and, presumably, statistical analysis of the metadata associated
with recent deliveries of CBDC into private wallets, such as times and
locations of the transactions, and the identities of the persons whose bank
accounts had been debited.  However, a strong degree of anonymity is necessary
for privacy.  For users of digital currency to trust the system, they must have
a way to verify that their activities will not be profiled. It is a human right
to have the choice to engage in transactions without being profiled, and this
flows from the right to privacy.

\subsection{Accessibility}

The universality of cash is made possible by the fact that it is accessible and
usable by almost everyone, even without the use of the Internet.  The authors
suggest that e-CNY would allow offline transactions.  We are sceptical of this
claim.  Specifically, we recognise that it is not possible to conduct a fair
exchange between two actors without a trusted third party~\cite{pagnia1999}.
The analogue equivalent of fair exchange is possible with physical cash as the
result of the fact that once physical tokens have been sent to the payee, they
cannot be unilaterally re-taken by the payer, as the transaction is settled on
receipt by way of exchange of possession.  With digital transactions, there is
not yet a way to prove to a recipient of tokens that the payer has not and will
not spend them a second time (``double-spending'') without indexing some third
party actor or system for record-keeping and verification.  One approach to
bridging the gap is to employ `trusted computing': a combination of hardware
and software that serves a master other than the user of a device.  However, we
note that trusted computing relies upon third-party tools and for this reason
cannot be implemented with complete equanimity for all parties doing the
trusting, and there will remain risks and costs to each of the parties given
the reliance on third-party tools.

So, how can we prevent double-spending in an offline context?  If the
definition of `offline' is framed in a way that includes third-party actors,
such as local observers, then it is certainly possible to achieve fair
exchange.  We refer to such scenarios as \textit{disconnected} rather than
fully offline.  Disconnected transactions can take place with local third-party
trust, perhaps via a local authority tasked with enforcing transaction
discipline, or via a separate device or process operated by the government.  It
might also be possible to have fully offline transactions if money is put in
escrow beforehand and then later revealed according to some rules for
redemption.  But we would argue that these transaction scenarios are not
offline in the same way that cash is, and for this reason, we believe that
digital currency, including e-CNY, would be an incomplete substitute with
limited accessibility for its users. Thus, the motivating reasons and factors
for the use of cash in certain contexts is not entirely transferable to those
of a CBDC, making any given CBDC a partial substitute at best.

Certainly, CBDC may help with certain aspects of inclusiveness, for example
with certain groups of remote areas that have great connectivity but are not
'plugged into' the digital economy.  However, seeing CBDC as a complete
substitute for physical cash makes it easy to justify getting rid of cash and
assuming that the policy objective of ensuring everyone's access to the economy
can be achieved through CBDC.  It cannot. This is especially risky where the solution being considered is one that eliminates the important privacy-preserving properties of cash. 

Cash must remain, for those that are
disabled, for those without access to banking facilities or money services
businesses, for those who are remotely located, for those who lack good network
connectivity or appropriate infrastructure, and because cash offers a private,
robust option to transfer physical objects with an agreed-upon value.  These
considerations are particularly important where the design of a CBDC system is
assumed to require the CBDC to be associated with a third party custodian or
other gatekeeper (for example, a bank or payment business) before a user can
make a payment, which results in exposing individuals to profiling and
discrimination based on their transactions.  For the avoidance of doubt, even
if central banks were to follow our advice, cash should still remain because
cash and CBDC are not perfect substitutes, as we also state in
Section~\ref{ss:use}. Thus, ensuring the continued availability for cash, and the existence of a cash-like digital payment instrument is both paramount, and aligned with the policy imperatives of central banks providing unfettered access to the economy, physical and digital in the same way they have enjoyed, or better, not worse.

\subsection{Regulatory Challenges}

The report's authors argue that the privacy problem is resolved with anonymous
accounts, wherein users reveal nothing.  Anonymous accounts often contravene
FATF-GAFI recommendations, and, as noted above, they are not really private
because they implicitly link their transactions to each other.  In many
countries in the developed world, custodians are subject to stringent rules
limiting the size of anonymous accounts that they manage on behalf of
customers.  Such limits are small relative to the size of transactions that can
take place anonymously using cash.

Although we assume that by reference to transactions of ``small value'', the
authors mean to suggest that anonymous accounts would include allowances made
for quasi-accounts such as pre-paid debit cards in other countries. We are
sceptical that this could include transactions of size comparable to the size
limits for anonymous cash transactions in other developed countries.  Please
note that pre-paid debit cards have stringent limits in much of the developed
world.  In the Eurozone, pre-paid debit cards are limited to a maximum size of
\euro150~\cite{eu5amld}.

We note that with payments made from non-custodial wallets, the regulatory
burden would be handled entirely by the receiving side, with the payer
providing no identifying information.  But the payer must have originally
received the digital currency from a regulated money services business (whether
a bank, payments business, or whatever type of regulated licenced entity in
that jurisdiction), and presumably, that money services business would have
already carried out appropriate checks for compliance with AML/KYC regulations.
This is analogous to the user engagement lifecycle for cash.  The bank knows
that Alice withdrew cash, and the recipient's bank knows that Bob deposited
cash, but nobody except Alice knows that she transacted with Bob.

\section{Common Misconceptions}

We suggest that at least three logical mistakes underpin the reasoning offered
by the PBOC. These three misconceptions have been adopted by many central banks
following the PBOC design, as well as by others that are evaluating approaches
to implementing and deploying CBDC.

The first assumption is that legal process is a sufficient barrier to protect
authorities (and others, such as private platform providers and attackers) from
unilaterally linking individual persons to their transactions.  We can
appreciate how much of the dialogue on the necessity of backdoors might lead
some authorities to believe that there is global agreement that it is
acceptable for authorities and other powerful groups to be able to unilaterally
gain access to whatever information they like, as long as they engage in a
specific bit of theatre.

This view has been rightly criticised over the years in the context of
cryptography~\cite{abelson2015,benaloh2018}, although it persists.  For
example, in the European Union, the eIDAS regulation~\cite{eidas} implicitly
suggests that formal procedures (law) for data protection by state-approved
``qualified trust providers'' is sufficient to protect the interests of
consumers and to keep corruption at bay.  But it is a dangerous fallacy to use
controversial regulations, or maximalist interpretations of FATF-GAFI
recommendations, as a straw man, or to suggest that, under such regimes,
powerful actors ``cannot'' access the information they want when they want to
do so.  In fact, they can access the information, whether or not doing so is
legal.  For an example of how this might play out in practice, consider the
increasingly common practice of ``parallel construction'' by law enforcement
organisations: the use of \textit{unlawful methods to gather secret evidence}
before constructing an argument on the basis of legally-obtained
evidence~\cite{hrw-parallel}.  Separately, it is also worth noting that access
to transaction information associated with an account reveals the identity of
its user even if the provided identity information is not known, so the
distinction among the tiers is vacuous at best.

The second assumption is that it is sufficient to compare the features of the
proposed wallets to the features of existing electronic payment accounts.
Essentially, this is an argument that people are already being tracked when
they make electronic payments all the time, which we can agree is true, so
therefore the new system introduces no new harm.  Unfortunately, this argument
ignores the fact that today, people have the option to make payments using
cash, and that part of the explicit motivation for the proposed wallet system
is to provide a public payment option that would substitute for cash in at
least some cases.  If achieved, such an effect would be significant whether or
not cash is completely eliminated.  Cash is already under significant strain in
China as well as many countries worldwide, wherein merchants struggle to accept
cash, given its high fixed costs and the low marginal revenues associated with
supporting cash infrastructure.  If consumers are given yet another option that
reduces cash use, then this effect would only be amplified, and governments
would soon come under pressure to stop requiring merchants to accept cash.
Note that in many countries, merchants are already not required to accept cash.
As a result, more transactions would be subject to the gaze of unilateral
surveillance. 

Instead, it should be presumed that central banks and governments would provide
their citizens with unfettered access to the economy, which is both physical
and, now, digital as well. Thus, access to the economy ought to be ensured both
in the physical realm, through cash, and in the digital realm, through digital
cash.  Crucially, this access must be on the same terms and in the same manner
as it has always been enjoyed.  Otherwise, users of money will experience a
reduction of their choices and the extent of their access to the economy,
exposing them to new harms, risks, and costs.

The third assumption is that wallets are really accounts that are provided or
registered by an authority as a precondition for their use.  This assumption
implies that authorities can control who can make payments, and, by extension,
that the assets contained in wallets are really owned by the authority (either
government or system operators), not by the individuals who are using the
assets to make payments.  To assume this is to entirely overlook non-custodial
wallets and systems in which users can use wallets that they trust, without
being forced to choose from among a set of solutions that give authorities what
they want.  It is possible to rely upon protocols rather than trusted
intermediaries to carry messages and make transactions; this has been the
explicit purpose of the greater part of a century of communications research,
whose foundations have brought us the Internet and modern e-commerce.  We do
not deny that governments might want to manufacture some hardware devices, for
example on behalf of certain government employees or to support groups who
cannot afford devices of their own, but that does not imply that all devices
must be registered or must run on trusted hardware.  We also do not deny that
governments might want to promulgate standards for such devices and recognise
certain auditors as legitimate certifiers of compliance with those standards,
but that does not imply that everyone must trust those auditors in particular
if they seek to interact with the system.

\section{Recommendations}

First and foremost, we should accept that a wallet is an `application to
generate, manage, or use private and public keys' that `can be implemented as a
software or hardware module'~\cite{iso-22739}.  We believe that wallets should
generally be brought by the consumers themselves and should generally not be
provided by financial institutions, asset custodians, technology platforms, or
other third parties, including commercial banks and other ``authorised
operators''.  We suggest that the PBOC and central banks around the world
should explicitly recognise non-custodial wallets, subject to the following six
recommendations for a CBDC in the digital economy:

\begin{enumerate}

\item A non-custodial wallet is a wallet that is under the direct control of
the consumer that owns the assets that it contains.  A non-custodial wallet is
not an account or quasi-account; it is not provided or administered by a third
party.

\item A consumer can have an arbitrary number of non-custodial wallets, and a
non-custodial wallet does not forcibly contain any identifying information that
can be used to link the assets it contains to the owner of the assets.

\item A consumer can withdraw digital currency from a regulated money or
payment services business into a non-custodial wallet using privacy-enhancing
technology, such as blind signatures or zero-knowledge proofs, to ensure that
the assets contained in the non-custodial wallet are fungible and not
distinguishable or recognisable by the regulated money services business or any
other parties as having been associated with the transaction in which the
digital currency was withdrawn.

\item The transactions of a central bank digital currency system would be
performed by regulated money services businesses.

\item A central bank digital currency system should include a mechanism to
ensure that its system operators do not equivocate.  This mechanism could be a
distributed ledger, for example, as we have proposed
elsewhere~\cite{goodell2021,goodell2019}.

\end{enumerate}

\section{Conclusion}

In this analysis, we have identified a set of assumptions and potential
concerns with the report by the PBOC, many of which are common among central
banks globally that are adopting the PBOC design, which is arguably the most
advanced CBDC design in the world at present.  These include the conflation of
wallets with accounts and generally employing an approach of data protection
rather than privacy by design in the proposed architecture for CBDC.  At the
same time, the approach taken by the PBOC is laudable in many ways, including
the depth with which it has developed its guiding principles which, we argue,
should also be adopted by other central banks.  We note that those principles
might be interpreted in a manner such that some principles have conflicts with
others or system-level implications that require attention.  We suggest that
central banks around the world are looking to the PBOC for guidance from a
design standpoint.  We argue that a balancing act is not needed and that all
the principles can be achieved, and this can only be done if we take the users
of CBDC and their privacy seriously.

It may be thought, as history has time and time again demonstrated to us, that
strict compliance with existing regulations must prevail, or that the most
powerful actors must benefit at the expense of the general public.  Not so: The
core message of our analysis is that a balancing act between regulatory
compliance, efficiency, privacy and inclusiveness should not take place.
Instead, policymakers should identify CBDC for its most essential use, as a
substitute for cash in specific instances for certain activities, whilst
interpreting the principles articulated by the PBOC in a way that does not lead
to contravening the rights that users have long been afforded through the use
of cash.  We offer our several specific recommendations in service of this end,
and we hope that the PBOC, and other central banks around the world, will
consider them as they seek to develop CBDC as a way to promote a
well-functioning digital economy that serves the public interest.

\section*{Acknowledgements}

We thank Professor Tomaso Aste for his continued support of our ongoing work
and Dann Toliver for inspiring our analysis.  We acknowledge the UCL Centre for
Blockchain Technologies and the Systemic Risk Centre at the London School of
Economics.


\begin{thebibliography}{1}\raggedright
\footnotesize

\bibitem{pboc2021}{
    Working Group on E-CNY Research and Development of the People's Bank of China.
    ``Progress of Research \& Development of E-CNY in China.''
    July 2021.
    [online]
    \url{http://www.pbc.gov.cn/en/3688110/3688172/4157443/4293696/2021071614584691871.pdf}
    [retreived 2021-07-16]
}
\bibitem{bloomberg2021}{
    Bloomberg News.
    ``China’s Digital Yuan Trial Reaches \$5.3 Billion in Transactions.''
    2021-07-16.
    [online]
    \url{https://www.bloomberg.com/news/articles/2021-07-16/china-s-digital-yuan-trial-reaches-5-3-billion-in-transactions}
    [retrieved 2021-08-10]
}
\bibitem{ripple2015}{
    Team Ripple.
    ``BIS Describes Peak Correspondent Banking
    [online]
    \url{https://ripple.com/insights/bis-describes-peak-correspondent-banking/}
    [retreived 2021-07-19]
}
\bibitem{sennero2019}{
    J Sennero.
    ``Swedish banks told to safeguard cash access despite rise in digital payments.''
    Reuters,
    2018-06-11.
    [online]
    \url{https://web.archive.org/web/20190714080523/https://www.reuters.com/article/us-sweden-banks-cash-idUSKBN1J70YY}
    [retreived 2021-07-19]
}
\bibitem{r3-cbdc}{
    G Calle and D Eidan.
    ``Central Bank Digital Currency: an innovation in payments.''
    R3 White Paper,
    April 2020.
    [online]
    \url{https://www.r3.com/reports/central-bank-digital-currency-an-innovation-in-payments/}
    [retrieved 2020-05-05]
}
\bibitem{world-bank}{
    The World Bank.
    ``Financial Inclusion: Overview.''
    2018-10-02.
    [online]
    \url{https://www.worldbank.org/en/topic/financialinclusion/overview}
    [retrieved 2021-07-15]
}
\bibitem{goldman2021}{
    D Goldman.
    ``China's Attempt to Avoid the American Tech Monopoly Trap.''
    American Affairs \textbf{5}(2), Summer 2021.
    [online]
    \url{https://americanaffairsjournal.org/2021/05/chinas-attempt-to-avoid-the-american-tech-monopoly-trap/}
    [retrieved 2021-07-19]
}
\bibitem{gladstein2021}{
    A Gladstein.
    ``Financial Freedom and Privacy in the Post-Cash World.''
    Cato Institute.
    Cato Journal,
    Spring/Summer 2021.
    [online]
    \url{https://www.cato.org/cato-journal/spring/summer-2021/financial-freedom-privacy-post-cash-world}
    [retrieved 2021-07-19]
}
\bibitem{khan2019}{
    V Khan and G Goodell.
    ``Libra: Is it Really About the Money?''
    August 2019.
    In \textit{Technology, Society, and Ethics},
    Jeremy Pitt, Ed.,
    ISBN: 979-8517435613,
    July 2021.
}
\bibitem{gorton2021}{
    G Gorton and J Zhang.
    ``Taming Wildcat Stablecoins.''
    Available at SSRN,
    2021-07-17.
    [online]
    \url{https://ssrn.com/abstract=3888752}
    [retrieved 2021-07-19]
}
\bibitem{boe2020}{
    Bank of England.
    ``Central Bank Digital Currency: opportunities, challenges and design.''
    Discussion paper,
    2020-03-12.
    [online]
    \url{https://www.bankofengland.co.uk/paper/2020/central-bank-digital-currency-opportunities-challenges-and-design-discussion-paper}
    [retrieved 2020-03-16]
}
\bibitem{fatf-recommendations}{
    Financial Action Task Force (FATF).
    \textit{The FATF Recommendations.}
    Updated February 2018.
    [online]
    \url{http://www.fatf-gafi.org/media/fatf/documents/recommendations/pdfs/FATF%20Recommendations%202012.pdf}
    [retrieved 2018-09-16]
}
\bibitem{iso-22739}{
    International Organization for Standardization (ISO).
    ``Blockchain and distributed ledger technologies -- Vocabulary.''
    \textit{ISO/22739:2020},
    First edition,
    2020.
    [online]
    \url{https://www.iso.org/obp/ui/#iso:std:iso:22739:ed-1:v1:en}
    [retrieved 2020-12-01]
}
\bibitem{un-human}{
    United Nations General Assembly.
    ``Universal Declaration of Human Rights.''
    217 (III) A. Paris, 1948.
    [online]
    \url{https://www.ohchr.org/EN/UDHR/Documents/UDHR_Translations/eng.pdf}
    [retrieved 2021-07-19]
}
\bibitem{fairfield2015}{
    J Fairfield and C Engel.
    ``Privacy as a Public Good.''
    Duke Law Journal \textbf{65}(385),
    2015.
    [online]
    \url{http://scholarship.law.duke.edu/dlj/vol65/iss3/1}
    [retrieved 2021-07-20]
}
\bibitem{goodell2020}{
    G Goodell, H Al-Nakib, and P Tasca.
    ``Digital Currency and Economic Crises: Helping States Respond.''
    LSE Systemic Risk Centre Special Papers SP 20,
    September 2020,
    Presented at 6th Annual Peer-to-Peer Financial Systems Workshop (P2PFISY 2020), December 2020.
    \url{https://systemicrisk.ac.uk/sites/default/files/2020-09/SP-20_0.pdf}
}
\bibitem{nissenbaum2017}{
    H Nissenbaum.
    ``Deregulating Collection: Must Privacy Give Way to Use Regulation?''
    May 2017.
    [online]
    \url{https://doi.org/10.2139/ssrn.3092282}
    [retrieved 2020-10-08]
}
\bibitem{goodell2021}{
    G Goodell, H Nakib, and P Tasca.
    ``A Digital Currency Architecture for Privacy and Owner-Custodianship.''
    \textit{Future Internet} 2021, 13(5), 130,
    May 2021.
    \url{https://doi.org/10.3390/fi13050130}
}
\bibitem{pagnia1999}{
    H Pagnia and F G\"artner,
    ``On the impossibility of fair exchange without a trusted third party.''
    Darmstadt University of Technology
    Department of Computer Science Technical Report TUD-BS-1999-02,
    1999.
    [online]
    \url{http://citeseerx.ist.psu.edu/viewdoc/download;jsessionid=9D84F245D05FE33F5B67E3BB4D1C9489?doi=10.1.1.44.7863&rep=rep1&type=pdf}
    [retrieved 2021-02-13]
}
\bibitem{eu5amld}{
    European Parliament.
    ``Directive (EU) 2018/843 of the European Parliament and of the Council of 30 May 2018.''
    2018-05-30.
    [online]
    \url{https://eur-lex.europa.eu/legal-content/en/TXT/?uri=CELEX:32018L0843}
    [retrieved 2018-09-26]
}
\bibitem{abelson2015}{
    H. Abelson, R. Anderson, S. Bellovin, J. Benaloh, M. Blaze, W. Diffie, J. Gilmore, M. Green, S. Landau, P. Neumann, R. Rivest, J. Schiller, B. Schneier, M. Specter, and D. Weitzner.
    ``Keys under doormats: mandating insecurity by requiring government access to all data and communications.''
    \textit{Journal of Cybersecurity} \textbf{1}(1),
    pp. 69--79,
    \texttt{doi:10.1093/cybsec/tyv009},
    2015-11-17.
    [online]
    \url{https://academiccommons.columbia.edu/doi/10.7916/D82N5D59/download}
    [retrieved 2019-03-11]
}
\bibitem{benaloh2018}{
    J. Benaloh.
    ``What if Responsible Encryption Back-Doors Were Possible?''
    Lawfare Blog,
    2018-11-29.
    [online]
    \url{https://www.lawfareblog.com/what-if-responsible-encryption-back-doors-were-possible}
    [retrieved 2018-12-11]
}
\bibitem{eidas}{
    The European Parliament and the Council of the European Union.
    ``Regulation (EU) No 910/2014 of the European Parliament and of the Council of 23 July 2014 on electronic identification and trust services for electronic transactions in the internal market and repealing Directive 1999/93/EC.''
    \textit{EUR-Lex}.
    2014-07-23.
    [online]
    \url{https://eur-lex.europa.eu/legal-content/EN/TXT/?uri=uriserv:OJ.L_.2014.257.01.0073.01.ENG}
    [retrieved 2021-10-25]
}
\bibitem{hrw-parallel}{
    S St Vincent.
    ``Dark Side: Secret Origins of Evidence in US Criminal Cases.''
    Human Rights Watch,
    2018-01-09.
    [online]
    \url{https://www.hrw.org/report/2018/01/09/dark-side/secret-origins-evidence-us-criminal-cases}
    [retrieved 2021-10-25]
}
\bibitem{goodell2019}{
    G Goodell and T Aste.
    ``Can Cryptocurrencies Preserve Privacy and Comply with Regulations?''
    \textit{Frontiers in Blockchain},
    May 2019.
    \url{https://doi.org/10.3389/fbloc.2019.00004}
}

\end{thebibliography}
\end{document}